\documentclass[twocolumn,showpacs,preprintnumbers,amsmath,amssymb]{revtex4}

\usepackage{graphicx}
\usepackage{dcolumn}
\usepackage{bm}

\begin{document}

\title{
Radio quiet neutron star 1E 1207.4-5209 as a source of
gravitational waves }

\author{Biping Gong}
\affiliation{ Department of Astronomy, Nanjing University, Nanjing
210093, P.R. China}

\begin{abstract}
There are four puzzles on 1E\,1207.4-5209: (1) the characteristic
age of the pulsar is much higher than the estimated age of the
supernova remnant; (2) the magnetic field inferred from spin-down
is significantly different from the value obtained from the
cyclotron absorption lines; (3) the spinning down of the pulsar is
non-monotonic; (4) the magnitude of the frequency's first
derivative varies significantly and its sign is also variable. The
third puzzle  can be explained by a wide binary system, with
orbital period from 0.2 to 6\,yr. This letter proposes that all
four puzzles can be explained naturally by  an ultra-compact
binary with orbital period of between 0.5 and 3.3\,min. 
With the shortest orbital period and a
close distance of 2\,kpc, the characteristic amplitude of
gravitational waves is $h\sim3\times10^{-21}$. It would be an
excellent source of gravitational-wave detectors such as the Laser 
Interferometer Space Antenna.

\end{abstract}
\draft 

\maketitle

\section{Introduction}

The radio quiet neutron star  1E\,1207.4-5209 (hereafter 1E1207)
is at the center of  the supernova remnant (SNR) PKS 1209$-$51/52.
It was discovered by Helfand $\&$ Becker~\cite{helfand84} with the
Einstein Observatory. The distance to the SNR is
$d=2.1^{+1.8}_{-0.8}$\,kpc~\cite{Giac00}. The X-ray spectrum of
the central source can be described by a thermal model which gives
 a distance of 2\,kpc~\cite{Mere96, Vas97, Zavlin98}.

The long observations devoted to 1E1207 both by Chandra and   by
XMM-Newton have unveiled a number of unique and somewhat
contradictory characteristics that, at the moment, defy standard
theoretical interpretations~\cite{Bignami04}.

The characteristic age of the pulsar, 200 to 900
\,kyr~\cite{Pavlov02}, is much larger than the estimated age of
the SNR, 3 to 20\,kyr~\cite{Roger88}.

The values of the spin-down luminosity, $\dot{E}\sim 1\times \rm
10^{34}\,erg\,s^{-1}$, and conventional magnetic field (B-field),
$B\sim 3\times 10^{12}$\,G, are typical for a radio
pulsars~\cite{Zavlin04}. However such a B-field is significantly
different from the value obtained from the cyclotron absorption
lines interpreted both in terms of electrons ($B\sim 8\times
10^{10}$\,G) as well as protons ($B\sim 1.6\times 10^{14}$\,G)~\cite{
Sanwal02, Luca04}.

Chandra and XMM-Newton observations indicate that the pulsar is
not spinning down steadily~\cite{Mere02, Bignami03, Zavlin04}.
Moreover, the  first derivative pulse frequency varies
significantly and its sign is also variable in different
observations~\cite{Zavlin04}.

The non-monotonic behavior of its pulse frequency, $\Delta\nu$, is
interpreted by three hypotheses: glitch, accretion and binary
(with an orbital period of 0.2 to 6\,yr), in which the binary hypothesis
is somewhat more plausible than the other two~\cite{Zavlin04}.



This letter analyzes that for binary pulsars with very small mass
function,  the Roemer time delay  in one orbital period cannot be
resolved.  Such pulsars  may thus be treated as ``isolated"
pulsars. However the pulse frequency and frequency derivatives of
such pulsars are still affected by the orbital motion at long time
scale, which causes anomalies like that of 1E1207.

Different from the orbital period of 0.2 to 6yr~\cite{Zavlin04},
the one that predicted by this letter is much shorter, 0.5 to
3.3\,min, therefore, 1E1207 is an ideal source of gravitational
waves.

\section{Orbital effect at long time scale}
Roemer time delay is the propagation time across the binary orbit,
which is given,
\begin{equation}
\label{delay} \frac{z}{c}=\frac{r\sin i}{c}\sin(\omega+f) \,,
\end{equation}
where $c$ is the speed of light, $r$ the distance between the
focus and the pulsar, $f$ the true anomaly, $\omega$ the angular
distance of the periastron from the node, and $i$ the orbit's
inclination. The orbital motion also causes the change of pulse
frequency, $\Delta\nu$,
\begin{equation}
\label{vnc}\frac{\Delta\nu}{\nu}=\frac{{\bf v}\cdot{\bf n}_{\rm
p}}{c}=K[\cos(\omega+f)+e\cos\omega] \,,
\end{equation}
where $K\equiv {2\pi a_{\rm p}\sin i}/{[cP_{\rm b}(1-e^2)^{1/2}]}$
is the semi-amplitude,
 $e$,
$P_{\rm b}$, $a_{\rm p}$ are eccentricity, orbital period, and
pulsar semi-major axis, respectively.

Small companion mass,  $i$ or $P_{\rm b}$ of a binary pulsar may
make the Roemer time delay of  Eq.~($\ref{delay}$) unmeasurable. A
binary pulsar may thus be treated as an ``isolated" pulsar.
Whereas, following calculation indicates that  Eq.~($\ref{delay}$)
and Eq.~($\ref{vnc}$) can still cause long-term effects on such
``isolated" pulsars.

For a binary pulsar,  the time received
by the observer (barycentric time) is, \vspace{-1mm}
\begin{equation}
\label{tb} t_{\rm b}=t_{\rm p}+\frac{z}{c} \,,
\end{equation}
where $t_{\rm p}$ is the proper time of the pulsar, and $z/c$ is
dependent on Kepler equation,
\begin{equation} \label{EM} E-e\sin
E=\bar{M}=\bar{n}t  \,,
\end{equation}
where $\bar{M}$, $E$ and $\bar{n}$ are mean anomaly, eccentric
anomaly and mean angular velocity, respectively. Note that $t$ is
the time of periastron passage, which is  uniform.

For a true isolated pulsar, we have $z/c=0$ in Eq.~($\ref{tb}$),
thus $t_{\rm b}=t_{\rm p}$, which means both $t_{\rm b}$ and
$t_{\rm p}$ are uniform. But for a binary pulsar system, $t_{\rm
b}$ is no longer uniform, whereas  $t_{\rm p}$ is still uniform.

Therefore, the proper time of the pulsar, $t_{\rm p}$, can be used
to replace the uniform time, $t$ of Eq.~($\ref{EM}$), then we have
$\bar{M}=\bar{n}t_{\rm p}$.

If  $\Delta\nu$ of  Eq.~($\ref{vnc}$) is averaged  over one orbit
period by the measured time, $t_{\rm b}$, then it gives
$$\left<\Delta \nu \right>
=\frac{1}{P_{\rm b}}\int^{P_{\rm b}}_{0}\Delta\nu dt_{\rm
b}=\frac{1}{P_{\rm b}}\int^{P_{\rm b}}_{0}\Delta\nu(dt_{\rm
p}+\frac{\dot{z}}{c}dt_{\rm p})$$ \vspace{-3mm}
$$
=\frac{1}{P_{\rm b}}\int^{P_{\rm
b}}_{0}\Delta\nu\frac{\dot{z}}{c}dt_{\rm p}=\frac{X}{P_{\rm
b}}\int^{P_{\rm b}}_{0}\Delta\nu \cos(\omega+E)\dot{E}dt_{\rm p}
$$
\vspace{-3mm}
\vspace{-3mm}
\begin{equation}\label{av1}
=\frac{XK\nu}{P_{\rm b}}\,\pi\,(1-\frac{e^2}{4})+O(e^4)  \,,
\end{equation}
where $X$ is the projected semi-major axis, $X\equiv a_{\rm p}\sin
i/c$.

In practical observation, an observer averages $\Delta\nu$ from 0
to $T$ ($T\gg P_{\rm b}$) through $t_{\rm b}$, the time received
by observer, without knowing the orbital period, $P_{\rm b}$, at
all. However if the pulsar measured is truly in a binary system,
$P_{\rm b}$ will affect the averaged result, as given in
Eq.~($\ref{av1}$), thus the averaged $\Delta\nu$ given by the
observer is
$$ \left<\Delta \nu\right> \ = \frac{1}{T}\int^{T}_{0}\Delta\nu
dt_{\rm b}\,=\,\frac{1}{T}\left(\int^{P_{\rm b}}_{0}\Delta\nu dt_{\rm
b}+\,...\, \right.$$
\vspace{-3mm}
\begin{equation} \label{av2}
\left. + \int^{P_{\rm b} N}_{P_{\rm b} (N-1)}\Delta\nu dt_{\rm
b}+\int^{T}_{P_{\rm b}N}\Delta\nu dt_{\rm b}\right)=\beta+
o(\beta\frac{P_{\rm b}}{T})
 \ ,
\end{equation}
where $\beta\equiv{XK\nu}\,\pi\,(1-{e^2}/{4})/{P_{\rm b}}$, and
$N$ is an integer.  Eq.~($\ref{av2}$) indicates that if a pulsar
is in a binary system, then $\left<\Delta \nu\right>$ measured by
the observer is actually contaminated by the long-term orbital
effect, $\beta$.

\section{Interpretation of  four puzzles and estimation  of orbital period of 1E\,1207}
 \subsection{Puzzle 1: pulsar age vs SNR age}
\label{sec:P1}
If the $\Delta\nu$ of Eq.~($\ref{av1}$) and
Eq.~($\ref{av2}$) (brackets $\left<,\right>$ are ignored
hereafter) are unchangeable then the effect actually  cannot be
measured. However $\beta$ contains orbital elements, $i$, $e$ and
$a$ (where $a$ is the semi-major axis of the orbit, $a=a_{\rm
p}M/M_2$, $M$ and $M_2$ are the total mass and companion mass
respectively, the mass of the pulsar is $M_1$) which are
long-periodic terms when the Spin-Orbit coupling effect is
considered. Therefore, $\Delta\nu$ is a function of time, and the
orbital effect induced $\dot{\nu}_{\rm L}$ can be found by
differentiating $\beta$ of Eq.~($\ref{av2}$), as given in detail
in the following subsection.

Thus the observational first derivative of the pulse frequency,
$\dot{\nu}_{\rm obs}$, is given by
\begin{equation} \label{dotpobs}
\dot{\nu}_{\rm obs}=\dot{\nu}+\dot{\nu}_{L}
\end{equation}
where $\dot{\nu}$ is the intrinsic one, which is caused by magnetic
dipole radiation. Thus the following relation can be obtained
\begin{equation} \label{nuobs} \frac{2\dot{\nu}_{\rm obs}}{\nu_{\rm obs}}
=(\frac{2\dot{\nu}}{\nu}+\frac{2\dot{\nu}_{L}}{\nu})\frac{\nu}{\nu_{\rm obs}}
\approx\frac{2\dot{\nu}}{\nu}+\frac{2\dot{\nu}_{L}}{\nu} \ \,.
\end{equation}
 Eq.~($\ref{nuobs}$) is actually
 \begin{equation}
\label{tau}
 -\frac{1}{\tau}=-\frac{1}{\tau_{\rm p}}+\frac{2\dot{\nu}_{L}}{\nu} \ \,
\end{equation}
where  $\tau=200-900$\,kyr is the  age corresponding to the
contaminated spin-down (by the long-term orbital effect).  In
other words, when the true age of the pulsar equals the age of
SNR, $\tau_{\rm p}=3-30$\,kyr is the true characteristic age of
the pulsar.

Putting $\tau$ and $\tau_{\rm p}$ into  Eq.~($\ref{tau}$), one
obtains two group solutions corresponding to maximum and minimum
magnitude of $\dot{\nu}_{L}$ and $\dot{\nu}$, respectively,
($\dot{\nu}_{L}=1.2 \times 10^{-11}\rm\, Hz~s^{-1}$,
$\dot{\nu}=-1.3 \times 10^{-11}\rm\, Hz~s^{-1}$); and
($\dot{\nu}_{L}=1.7\times 10^{-12}\rm\, Hz~s^{-1}$,
$\dot{\nu}=-1.9\times 10^{-12}\rm\,Hz~s^{-1}$).

This implies that the magnitude of $\dot{\nu}$ and $\dot{\nu}_{L}$
are much larger than that of $\dot{\nu}_{\rm obs}$, since
$\dot{\nu}_{L}$ and  $\dot{\nu}$ nearly cancel each other out.
Therefore the age puzzle, $\tau_{\rm p}\ll\tau$, can also be
explained.

\subsection{Puzzle 2: B-field}
Section \ref{sec:P1} shows that the measured $\dot{\nu}_{\rm obs}$
may under-estimate the true intrinsic pulse frequency derivative,
$\dot{\nu}$. Therefore, the B-field $3\times10^{12}$\,G inferred
from $\dot{\nu}_{\rm obs}$ may be under-estimated also.

The two $\dot{\nu}_{\rm L}$ obtained through  Eq.~($\ref{tau}$)
correspond to two magnetic dipole radiation-induced  $\dot{\nu}$,
and therefore, to two B-field, $3\times10^{13}$\,G and
$1\times10^{13}$\,G, respectively. It is easy for them to
reconcile with the high B-field option,
$B=1.6\times10^{14}$\,G~\cite{Sanwal02},
i.e., by  assuming the magnetic inclination
angle, $\alpha=11^\circ$ and  $\alpha=4^\circ$, respectively.

However, it is very difficult for these two B-fields to reconcile
with another option, $B=8\times10^{10}$\,G~\cite{Sanwal02}.
Therefore, the B-fields inferred from the true intrinsic spin-down
favors that 1E1207 is a magnetar.

\subsection{Puzzle 3: non-monotonic spin-down and estimation of orbital period}
In the gravitational two-body problem with spin, each body  precesses
in the gravitational field of its companion (geodetic precession),
with precession velocity of 1 Post-Newtonian order (PN)~\cite{bo}.
The Spin-Orbit coupling causes long-periodic variations in the six
orbital elements, $i$, $e$, $a$, $\bar{M}$, $\omega$ and $\Omega$
(longitude of the ascending node)~\cite{gong}. By the definition
of $K$ and $X$, $\beta$ of Eq.~($\ref{av2}$) can be rewritten as,
\begin{equation}
\label{a}{\Delta\nu}=\beta=\frac{GM\nu}{2\pi c^2 a}\,\rho \,,
\end{equation}
where $\rho\equiv\pi\sin^2 i\,(M_2/M)^2\,(1-e^2/4)/\sqrt{1-e^2}$.
According to Eq.~($\ref{a}$), $\beta$ contains the orbital elements,
$e$, $i$ and $a$, which are all long-periodic terms when the
Spin-Orbit effect is considered. However, the variation of $i$ is
much smaller ($S/L$ times, $S$ and $L$ are the spin and orbital
angular momentum respectively) than that of $a$ and
$e$~\cite{gong}. Thus the long-period variation of
Eq.~($\ref{av2}$)  can be written in a Taylor series as
\begin{equation}
\label{avnc2}{\Delta\nu}=\beta=\beta_0+\dot{\beta}t+...
=\beta_0-\beta\,\frac{\dot{a}}{a}\,(1-\xi)t+... \,,
\end{equation}
where
\begin{equation}
\label{xi}
\xi\equiv\frac{(1-e^2)e^2}{2(1+e^2)(1-e^2/4)}+\frac{e^2}{1+e^2}
\,,
\end{equation}
and
\begin{equation}
\label{avnc2a}\frac{\dot{a}}{a}=\frac{GL(1+e^2)}{c^2a^3(1-e^2)^{5/2}}\
(2+\frac{3M_2}{2M_1})\ (P_{y}Q_{x}-P_{x}Q_{y})
\end{equation}
where $P_{x}$, $P_{y}$, $Q_{x}$, $Q_{y}$ are sine and cosine
functions of $\omega$ and $\Omega$~\cite{gong}. The orbital period
$P_{\rm b}$ of a few minutes corresponds to $\dot{\omega}_{GR}\sim
10^{-5}\rm \,s^{-1}$, which corresponds to $\dot{a}/a\sim
10^{-6}\rm\,s^{-1}$. Thus in the observation time span, $\Delta
t\sim 10^2$\,ks, $a$ has changed like, one tenth of its period,
which means $\omega$ has changed by $\pi/5$. This actually
corresponds to a large variation amplitude in $\Delta a$.

Define  $\delta a\equiv|\Delta a/a|_{\max}$, then the maximum and
minimum $a$ of Eq.~($\ref{a}$) are $a_{\max}=a(1+\delta a)$, and
$a_{\min}=a(1-\delta a)$ respectively. The discrepancy in
$\Delta\nu$ corresponds to the error bar of each observation is
given,
\begin{equation}
\label{errorbar}\delta\nu=\frac{GM\nu\rho}{2\pi c^2a} \,
\left(\frac{1}{1-\delta a}-\frac{1}{1+\delta a}\right)
 =\beta\frac{2\delta a}{1-(\delta a)^2} \,.
\end{equation}
The fact that the amplitude of $\delta\nu$ is not much larger than
a few $\mu$Hz \cite{Zavlin04} demands that $a>|\Delta a|$. Thus
the maximum  $\delta a$ can only be like $\delta a=0.9$, whereas
$\delta a=1$ is not allowed.

From the point of view of Eq.~($\ref{a}$), both error bars in one
observation and discrepancy for different observations are
dependent of the variation of $a$. The difference is that the
discrepancy among different observations, i.e., Jan 2000, Aug
2002, corresponds to a much longer time scale, in which
$\Delta\nu$ is modulated by  both $\omega$ and $\Omega$ (the
period of $\Omega$ is comparable to that of   $\omega$) for many
periods. Whereas in one observation ($10^1$\,ks\,--\,$10^2$\,ks),
the time may be just enough for $\Delta\nu$ to vary in a few
periods of $\omega$, or even less than a period of $\omega$.


The jump of $\Delta\nu$ between Dec 2001 and Jan
2002~\cite{Zavlin04}, can be explained by the variation of
$\omega$ and $\Omega$, which causes relatively sharp variation in
$a$ and thus significant variation in  $\Delta\nu$.

The second term at the right hand side of Eq.~($\ref{avnc2}$)
actually corresponds to $\dot{\nu}_{L}$, which is given in
magnitude as
\begin{equation}
\label{eq2}
\dot{\nu}_{L}=\beta\,\frac{\dot{a}}{a}\,(1-\xi)
 \,.
\end{equation}
Putting the two $\dot{\nu}_{\rm L}$ obtained in Eq.~($\ref{tau}$)
into Eq.~($\ref{eq2}$), we have  two curves $\rho$ vs $P_{\rm b}$
corresponding to $\dot{\nu}_{\rm L1}$ and $\dot{\nu}_{\rm L2}$
respectively, as shown in Fig.~\ref{Fig:f1}.

Similarly putting $\Delta\nu_1=0.18\,\mu$Hz and
$\Delta\nu_2=4.2\,\mu$Hz into Eq.~($\ref{av2}$), we have two
$\rho$ vs $P_{\rm b}$ curves as shown in Fig.~\ref{Fig:f1}, which
correspond to minimum and maximum discrepancies in $\nu$ or 
error bars in different
observations of Zavlin et al~\cite{Zavlin04}.

The maximum orbital period, 3.3\,min, is given by the cross
section of $\Delta{\nu}_2$ and $\dot{\nu}_{\rm L2}$ at $C$ as
shown in Fig.~\ref{Fig:f1}. The minimum orbital period is 0.1\,min
corresponding to $A$ given by $\Delta{\nu}_1$ and $\dot{\nu}_{\rm
L1}$.

In the area ABCD of Fig.~\ref{Fig:f1}, a point, i.e., with $P_{\rm
b}=0.7$\,min and  $\rho=0.008$ can be found. Assuming
$M_1=1.4M_{\odot}$, $M_2=0.2M_{\odot}$ and $e=0$,
$\rho\approx0.09\sin^2 i$ is obtained, and by the definition of
$\rho$, $\sin i\approx0.3$ can be obtained. In turn $X$ is given,
$X\approx2.6$\,ms, which is smaller than the time resolution of
observation, 5.7\,ms or 2.9\,ms~\cite{Zavlin04}. Therefore, the
modulation induced by the orbital motion may not be detected from
the timing observation. This is consistent with the fact that the
side band corresponding to $P_{\rm b}$ of a few minutes has not
been found in 1E1207.

Therefore, the companion of 1E1207  should have low mass and be
compact enough. Because the undetected orbital  modulation implies
that $X$ must be small; and the short orbital period demands that
the companion be compact object like low-mass neutron
star~\cite{carr03} or strange star~\cite{xu03,xu05}. A  white
dwarf star companion is unlikely due to the separation of the two
stars is almost equal to the radius of a white dwarf star when the
orbital period is of 1\,min.

\subsection{Puzzle 4: magnitude and sign of $\dot{\nu}_{\rm obs}$}

The values of $P_{\rm b}$, $\rho$ and  the period of $\omega$
corresponding to the four points ABCD of Fig.~\ref{Fig:f1} are
shown in Table~\ref{tb:t1}.
\begin{table}
\begin{center}
\caption{\label{tb:t1} Results correspond to four points of
Fig.~\ref{Fig:f1} }
\begin{tabular}{cccr}
\\[-1.5ex]
\hline\hline\rule{0mm}{2.5ex} point &  $P_{\rm b}$(min)  & \
$\rho$ & \ \ $T_{\rm \omega}$(day) \
\\[0.5ex] \hline
\rule{0mm}{2.5ex}

A  & \ 0.1 \ \ & $6\times10^{-4}$   &  0.6  \\

B  & \ 0.3 \ \ & $1\times10^{-3}$  &  4  \\

C  & \ 3.3 \ \ & $3\times10^{-1}$   &  183  \\

D  & \ 1.0 \ \ & $1\times10^{-1}$  &  25  \\[0.5ex]

\hline \hline
\end{tabular}
\end{center}
{\small  $T_{\rm \omega}=2\pi a/\dot{a}$
 represents the period of $\dot{a}/a$.
}
\end{table}

As given by  Eq.~($\ref{a}$) and Eq.~($\ref{eq2}$), both
$\Delta\nu$ and $\dot{\nu}_{\rm L}$ vary with $a$ which is in turn
modulated by the period, $T_{\rm \omega}$ (and $T_{\rm \Omega}$)
as shown in Table~\ref{tb:t1}. In the case $P_{\rm b}=1$\,min, the
period of $T_{\rm \omega}$ is $\sim25$~days, therefore the period
of variation of $\dot{\nu}_{\rm L}$ is approximately $\sim25$~days
also (recall the period, $T_{\rm \Omega}$, is comparable with
$T_{\rm \omega}$).

On the other hand, the intrinsic $\dot{\nu}$ changes steadily,
which means $\dot{\nu}$ and $\dot{\nu}_{\rm L}$ some times
cancelling out, and some times  have the same sign and enhancing,
thus $\dot{\nu}_{\rm obs}$ can both be $\sim 10^{-14}\rm\,s^{-2}$
($\dot{\nu}$ and $\dot{\nu}_{\rm L}$ cancelled out); and $\sim
10^{-11}\rm\,s^{-2}$ ($\dot{\nu}$ and $\dot{\nu}_{\rm L}$ enhanced).
This well explains the observations of Zavlin et
al~\cite{Zavlin04}, which show that $\dot{\nu}_{\rm obs}$ can have
very different magnitude, $\sim 10^{-14}\rm\,s^{-2}$ and $\sim
10^{-11}\rm\,s^{-2}$ and its sign is also changeable at different
epochs.

Eq.~($\ref{a}$) and Eq.~($\ref{eq2}$) actually predict that
$\Delta\nu$ and $\dot{\nu}_{\rm L}$ can vary with periods of
 days, thus $\dot{\nu}_{\rm obs}$ can change sign in order
of days, or even during one observation ($10^2$\,ks). Comparing
Table~\ref{tb:t1} with observation may extract the period $T_{\rm
\omega}$ and $T_{\rm \Omega}$ and therefore determine the orbital
period $P_{\rm b}$.

\section{Discussion}
Therefore, all four puzzles can be explained naturally by an
ultra-compact binary system.


The best spectral model describes the  continuum as the  sum of
two blackbody curves with $kT=0.211\pm0.001$\,kev, for an emitting
radius $R=2.95\pm0.05$\,km; and  $kT=0.40\pm0.02$\,kev
($R=250\pm50$\,m)~\cite{Bignami03}. It is possible that these two
emitting radii are from the hot spot of two stars,  1E1207 and its
companion.



The characteristic amplitude of gravitational waves from a binary
system is ~\cite{Thorne}
$$h=1.4\!\times\!10^{-20}(\frac{\mu}{M_{\odot}})(\frac{M}{M_{\odot}})^{2/3}
(\frac{P_{\rm
b}}{1\rm\,hr})^{-2/3}(\frac{d}{100\rm\,pc})^{-1}f(e)$$ \vskip-7mm
\begin{equation}
\label{thorne}\sim3\times 10^{-21} \,,
\end{equation}
where $\mu$ is the reduced mass which equals $\mu=0.18M_{\odot}$
when $M_1=1.4M_{\odot}$ and $M_2=0.2M_{\odot}$; $d=2$\,kpc is the
distance; $P_{\rm b}=1$\,min is orbital period; and $f(e)$ is
given by
$f(e)=(1+\frac{73}{24}e^2+\frac{37}{96}e^{4})/(1-e^2)^{7/2}$,
which is assumed $f(e)\approx1$. In such case,
$GM/(c^2a)\sim (v/c)^2\sim9\times10^{-5}$ ($v$ is a characteristic
orbital velocity), which means equations,
Eq.~($\ref{avnc2}$)--Eq.~($\ref{eq2}$), based on the
Post-Newtonian approximation are good enough to describe the
dynamics of the ultra-compact binary system.

The time scale of  coalescing of 1E1207 corresponding to $P_{\rm
b}= 0.5$\,min is $\sim27$\,yr. In order to be consistent with the
fact that it was discovered in 1984 and is still there, it is
necessary that $P_{\rm b}\geq0.5$\,min. Therefore the most
probable orbital period of 1E1207 is (0.5--3.3)\,min.

The low wave frequency, $\sim10^{-2}$\,Hz, and the extremely large
wave amplitude means that 1E1207 is an ideal source for the space detector
Laser Interferometer Space Antenna.

\begin{acknowledgments}
I thank A. R\"udiger and T. Kiang for very helpful comments and
corrections on the presentation of this letter. I thank T.Y. Huang
and X.S. Wan for useful comments on the mathematics in this
letter.

\end{acknowledgments}

\clearpage

\begin{figure}[h]

\includegraphics[87,87][500,700]{figure1.eps}
 \caption{\label{Fig:f1} The two $\Delta\nu$ curves ($\Delta\nu_1=0.18\,\mu$Hz,
and $\Delta\nu_2=4.2\,\mu$Hz) intersect with two $\dot{\nu}_{\rm
L}$ curves ($\dot{\nu}_{\rm L1}=1.2 \times 10^{-11}\rm\,Hz~s^{-1}$
and $\dot{\nu}_{\rm L2}=1.7\times 10^{-12}\rm\,Hz~s^{-1}$) at four
points ($A$, $B$, $C$ and $D$). Therefore, the binary parameters
of 1E1207 are contained in the area $ABCD$, in which the minimum
and maximum values of $\rho$ ($=
{\pi}\sin^2i(M_2/M)^2(1-e^2/4)/\sqrt{1-e^2}$) are from
$4\times10^{-4}$ to $2\times10^{-1}$, and those of $P_{\rm b}$ are
from 0.1\,min to 3.3\,min. Considering the coalescing time
constraint, the minimum $P_{\rm b}$ should be 0.5\,min. The
vertical arrow represents $P_{\rm b}=0.5$\,min.}
\end{figure}

\end{document}